\documentclass[aps,superscriptaddress,twocolumn,showpacs]{revtex4}
\usepackage{amsmath}
\usepackage{graphicx}
\begin{document}
\title{Relativistic X-Ray Free Electron Lasers in the Quantum Regime}
\author{Bengt Eliasson}
\affiliation{Institut f\"ur Theoretische Physik, Fakult\"at f\"ur Physik und Astronomie,
Ruhr--Universit\"at Bochum, D-44780 Bochum, Germany}
\author{P. K. Shukla}
\affiliation{International Centre for Advanced Studies in Physical Sciences \& Institute for Theoretical Physics,
Fakult\"at f\"ur Physik und Astronomie, Ruhr--Universit\"at Bochum, D-44780 Bochum, Germany}
\affiliation{Department of Mechanical and Aerospace Engineering \& Center for Energy Research, University of 
California San Diego, La Jolla, CA 92093, U. S. A.}
\received{15 December 2011}
\revised{28 February 2012}
\begin{abstract}
We present a nonlinear theory for relativistic X-ray free electron lasers in the quantum regime, using a
collective Klein-Gordon (KG) equation (for relativistic electrons), which is coupled with the Maxwell-Poisson 
equations for the electromagnetic and electrostatic fields. In our model, an intense electromagnetic wave 
is used as a wiggler which interacts with a relativistic electron beam to produce coherent tunable radiation. 
The KG-Maxwell-Poisson model is used to derive a general nonlinear dispersion relation for parametric 
instabilities in three-space-dimensions, including an arbitrarily large amplitude electromagnetic wiggler field. 
The nonlinear dispersion relation reveals the importance of quantum recoil effects and oblique scattering of the 
radiation that can be tuned by varying the beam energy.
\end{abstract}

\maketitle

With the recent development of X-ray free-electron lasers (FELs) \cite{Hand09} there
are new possibilities to explore matter on atomic and single molecule
levels. On these length scales, of the order of a few {\AA}ngstr\"oms, quantum
effects play an important role in the dynamics of the electrons. Quantum effects 
have been measured experimentally both in the degenerate electron gas in metals 
and in warm dense matter \cite{Glenzer}, and must also be taken into account in intense laser-solid density plasma
interaction experiments \cite{Andreev}. The theory of the FEL was originally developed in 
the framework of quantum mechanics \cite{Madey71}, but where Planck's constant cancelled 
out in the final FEL gain formula producing a classical result. It was subsequently shown that 
classical theory can be used and quantum effects can be neglected \cite{Hopf76},
if the photon momentum recoil is not larger than the beam momentum spread 
\cite{Bonifacio85,Schroeder01,Bonifacio05,Bonifacio05b}. To overcome the technical limitation 
on the wiggler wavelength for a static magnetic field wiggler, it was suggested that it can be 
replaced by an electromagnetic (EM) wiggler or by a  plasma wave wiggler \cite{Yan86} to generate 
short wavelength radiation. In such a situation, it turns out that quantum recoil effects can be important.
The Klein-Gordon equation (KGE) for a single electron was used to derive a general set of quantum 
mechanical  equations for the FEL \cite{McIver79}, while a single-electron Schr\"odinger-like equation
for the dynamics of the FEL was derived using the field theory \cite{Preparata88}.  Furthermore, by using 
a multi-electron FEL Hamiltonian, it was shown that quantum effects can lead to the  splitting of the 
radiated spectrum into narrow bands for short  electron bunches \cite{Bonifacio94,Bonifacio08}.
Relativistic and collective quantum effects have been studied for FELs using
Wigner \cite{Serbeto08,Piovella08} and quantum fluid \cite{Serbeto09} models.

In this Letter, we shall use a collective KGE to derive a dynamic model for the quantum free electron laser.
In our model, we assume that the wave function $\psi$ represents an ensemble of electrons, so that
the resulting charge and current densities act as sources \cite{Takabayasi53} for the self-consistent 
electrodynamic vector and scalar potentials ${\bf A}$ and $\phi$, respectively.

The KGE in the presence of the electrodynamic fields reads
\begin{equation}
{\cal W}^2\psi-c^2{\cal P}^2\psi-m_e^2c^4\psi=0,
\label{KG}
\end{equation}
where the energy and momentum operators are ${\cal W}=i\hbar{\partial}/{\partial t}+e\phi$, and 
${\cal P}=-i\hbar\nabla+e{\bf A}$, respectively. Here $\hbar$ is Planck's constant divided by $2\pi$, $e$  the magnitude of 
the electron charge, and $m_e$ the electron rest mass, and $c$ the speed of light in vacuum. 
The electrodynamic potentials are obtained self-consistently from the Maxwell equations
\begin{equation}
\frac{\partial^2{\bf A}}{\partial t^2}+ c^2 \nabla\times(\nabla\times{\bf A})
+  \nabla\frac{\partial \phi}{\partial t}=\mu_0 c^2{\bf j}_e,
\label{wave3}
\end{equation}
and
\begin{equation}
   \nabla^2\phi+\nabla\cdot\frac{\partial {\bf A}}{\partial t}
=-\frac{1}{\varepsilon_0}(\rho_e+\rho_i),
   \label{Poisson}
\end{equation}
where $\mu_0$ is the magnetic permeability, 
$\varepsilon_0=1/\mu_0 c^2$ the electric permittivity in vacuum, and $\rho_i= e n_0$ 
a neutralizing positive charge density due to ions,  where $n_0$ is the equilibrium electron number
density.  The electric charge and current densities of the electrons are
$\rho_e=-e\left[\psi^\ast {\cal W}\psi+\psi({\cal W}\psi)^\ast\right]/2 m_e c^2$, and
${\bf j}_e=- e\left[\psi^\ast{\cal P}\psi+\psi({\cal P}\psi)^\ast\right]/2 m_e$,
respectively. They obey the continuity equation ${\partial \rho_e}/{\partial t}+\nabla\cdot{\bf j}_e=0$.

It is convenient to carry out the calculations in the beam frame, and then to Lorentz transform the
result into the laboratory frame. We assume that the beam is propagating along the $z$-axis, in the 
opposite direction of the laser wiggler beam.  For the laser wiggler field, we consider for simplicity 
a circularly polarized EM wave of the form ${\bf A}_0=(1/2)\widehat{\bf A}_0\exp(-i\omega_0 t+i{\bf k}_0\cdot{\bf r})$+ complex conjugate, 
with $\widehat{\bf A}_0=(\widehat{\bf x}+i\widehat{\bf y})\widehat{A}_0$,
where $\omega_0$ is the laser wave frequency and ${\bf k}_0=k_0\widehat{\bf z}$ the wave vector, and
$\widehat{\bf x}$, $\widehat{\bf y}$ and $\widehat{\bf z}$ the unit vectors in the $x$, $y$ and $z$ directions,
respectively.  Due to the circular polarization, the oscillatory parts in the nonlinear term proportional to $A^2$ 
in the KGE vanish. Hence, our starting point is the nonlinear dispersion relation in the beam frame, with primed 
variables,  where the plasma is at rest. It reads \cite{Eliasson11}
\begin{equation}
\! 1-\frac{(\omega_{pe}')^2}{4 \gamma_A^3 m_e^2 c^2}\frac{D_A'(\Omega',{\bf K}')}
{D_L'(\Omega',{\bf K}')}
\sum_{+,-}\frac{e^2|{\bf k}_\pm'\times\widehat{\bf A}_0'|^2}{(k_{\pm}')^2 D_A'(\omega_{\pm}',{\bf k}_{\pm}')}=0,
\label{FEL_disp}
\end{equation}
where the electron plasma oscillations in the presence of the EM field are represented by
\begin{equation}
\begin{split}
& D_L'(\Omega',{\bf K}')= \frac{(\omega_{pe}')^2}{\gamma_A}-(\Omega')^2
\\
&+\frac{\hbar^2 [c^2 (K')^2-(\Omega')^2]} {4\gamma_A^2 m_e^2 c^4} D_A'(\Omega',{\bf K}').
\end{split}
\end{equation}
Here $\Omega'$ and ${\bf K}'$ are the frequency and wave vector of the plasma oscillations, respectively, 
$\gamma_A=(1+a_0^2 )^{1/2}$ is the relativistic gamma factor due to the large amplitude EM field, and 
$a_0=e |\widehat{A}_0'|/m_e c$ is the normalized amplitude of the EM wave (the wiggler parameter).
The dispersion relation for the beam oscillations in the presence of
a large amplitude EM wave is given by $D_L'(\Omega',{\bf K}')=0$. The EM sidebands are governed by
\begin{equation}
  D_A'(\omega_\pm',{\bf k}_\pm')=c^2 (k_\pm')^2-(\omega_\pm')^2+(\omega_{pe}')^2/\gamma_A,
\end{equation}
where $\omega_{\pm}'=\omega_0'\pm \Omega'$ and ${\bf k}_{\pm}'={\bf k}_0'\pm{\bf K}'$, 
and $\omega_0'$ and $k_0'$ are related through the nonlinear dispersion relation 
$(\omega_0')^2=c^2 (k_0')^2+(\omega_{pe}')^2/\gamma_A$. We also denoted 
$D_A'(\Omega,{\bf K})=c^2 (K')^2-(\Omega')^2+(\omega_{pe}')^2/\gamma_A$.
We have neglected the two-plasmon decay \cite{Drake74}, which would give rise to terms proportional to
$|{\bf k}_\pm'\cdot\widehat{\bf A}_0'|^2$ in Eq. (\ref{FEL_disp}).

To move from the beam frame to the laboratory frame, the time and space variables are Lorentz transformed
as $t'=\gamma_0 (t-v_0 z/c^2)$, $x'=x$, $y'=y$ and $z'=\gamma_0(z-v_0 t)$, where ${\bf v}_0=v_0 \widehat{\bf z}$
is the beam velocity, and $\gamma_0=1/\sqrt{1-v_0^2/c^2}$ the gamma factor due to the relativistic beam speed.
The corresponding frequency and wavenumber transformations are  thus $\omega'=\gamma_0(\omega-v_0 k_z)$,
$k_x'=k_x$, $k_y'=k_y$, and $k_z'=\gamma_0(k_z-{v_0 \omega}/{c^2})$, They apply to the frequency 
and wave vector pairs ($\Omega$, ${\bf K}$), ($\omega_0$, ${\bf k}_0$) and ($\omega_\pm$, ${\bf k}_\pm$). The plasma 
frequency is transformed as $\omega_{pe}'=\omega_{pe}\sqrt{\gamma_A/\gamma}$, where 
$\gamma=(1+p_0^2/m_e^2 c^2+ a_0^2)^{1/2}$ is the total gamma factor and ${\bf p}_0=\gamma m_e {\bf v}_0$ the
relativistic electron momentum. Since the components of $\widehat{\bf A}_0$ are perpendicular to the beam velocity direction, 
they are not affected by the Lorentz transformation, hence $\widehat{\bf A}_0'=\widehat{\bf A}_0$. We also use the relation $\gamma_A\gamma_0=\gamma$, and observe that the expressions of the form $(\omega')^2-c^2 (k')^2=\omega^2-c^2 k^2$
are Lorentz invariant. This yields $D_L'(\Omega',{\bf K}')=\gamma_0^2 D_L(\Omega,{\bf K})$, with
\begin{equation}
\begin{split}
  & D_L(\Omega,{\bf K})=
\frac{\omega_{pe}^2\gamma_A^2}{\gamma^3}-(\Omega-v_0 K_z)^2
\\
&+\frac{\hbar^2 (c^2 K^2-\Omega^2)}
{4\gamma^2 m_e^2 c^4} D_A(\Omega,{\bf K}),
\end{split}
\label{DL3}
\end{equation}
and $D_A'(\omega_\pm',{\bf k}_\pm')=D_A(\omega_\pm,{\bf k}_\pm)\equiv c^2 k_\pm^2-\omega_\pm^2+\omega_{pe}^2/\gamma$,
and, similarly, $D_A'(\Omega',{\bf K}')=D_A(\Omega,{\bf K})$. In the laboratory frame,  Eq. (\ref{FEL_disp}) is of the form
\begin{equation}
\begin{split}
&1-\frac{\omega_{pe}^2}{4 \gamma^3 m_e^2 c^2}\frac{D_A(\Omega,{\bf K})}
{D_L(\Omega,{\bf K})}\sum_{+,-}\frac{e^2|{\bf k}_\pm'\times\widehat{\bf A}_0|^2}{(k_{\pm}')^2 D_A(\omega_{\pm},{\bf k}_{\pm})}=0.
\end{split}
\label{FEL_disp2}
\end{equation}
Using ${\bf K}=K_x\widehat{\bf x}+K_y\widehat {\bf y}+K_z\widehat {\bf z}$, we have $K^2=K_z^2+K_\perp^2$ with $K_\perp^2=K_x^2+K_y^2$, so that 
$|{\bf k}_\pm'\times \widehat{\bf A}_0|^2=\{2 \gamma_0^2[k_0\pm K_z+(v_0/c^2)(\omega_0\pm\Omega)]^2+K_\perp^2\}|\widehat{A}_0|^2$,
and $(k_\pm')^2=\gamma_0^2[k_0\pm K_z+(v_0/c^2)(\omega_0\pm\Omega)]^2+K_\perp^2$. 

For the resonant backscattering instability, we have $|D_A(\omega_+,{\bf k}_+)|\gg|D_A(\omega_-,{\bf k}_-)|$.
Also, for $\Omega\approx v_0 K_z$, $v_0\approx -c$, and $\gamma_0 K_z \gg k_0,\, K_\perp$, we have
$|{\bf k}_\pm'\times \widehat{\bf A}_0|^2/(k_\pm')^2 \approx 2 |\widehat{A}_0|^2$. In this limit, Eq. (\ref{FEL_disp2}) is written as
\begin{equation}
  D_L(\Omega,{\bf K}) D_A(\omega_{-},{\bf k}_{-})= \frac{\omega_{pe}^2 D_A(\Omega,{\bf K})}{2\gamma^3} a_0^2.
  \label{coupling}
\end{equation}
By using $\Omega\approx v_0 K_z$, the expression for $D_L(\Omega,{\bf K})$ can be simplified as
\begin{equation}
   D_L(\Omega, {\bf K})=  {\Omega_{p}^2}-(\Omega-v_0 K_z)^2,
   \label{DL2}
\end{equation}
where
\begin{equation}
  \Omega_{p}=\sqrt{\frac{\omega_{pe}^2\gamma_A^2}{\gamma^3}+\frac{\hbar^2 (K_z^2\gamma_A^2+K_\perp^2\gamma^2)^2}{4\gamma^6 m_e^2}}.
  \label{Ohm_pm}
\end{equation}
Equation (\ref{DL2}) is valid for  $\hbar\omega_{pe}\gamma_A/\gamma^{3/2} m_e c^2\ll 1$ (which is always fulfilled), and
$\omega_{pe}\gamma_A/|v_0|\gamma^{3/2}\ll K_z \ll m_e |v_0|\gamma^{3/2}/\hbar\gamma_A$. The latter condition, 
with $|v_0| \approx c$, gives $\gamma_A/\lambda_e\gamma^{3/2} \ll K_z \ll \gamma^{3/2}/\lambda_C\gamma_A$, where $\lambda_e=c/\omega_{pe}$ is the electron skin depth and $\lambda_{C}= \hbar / m_e c$ the reduced Compton length.
We note that $\Omega_{p}$ contains a combination of the collective beam plasma oscillation and quantum recoil effects, 
which lead to a splitting of the beam mode into one slightly upshifted and one downshifted mode.

The resonant $\Omega$ and ${\bf K}$ are obtained by simultaneously setting $D_L(\Omega, {\bf K})=0$ and $D_A(\omega_{-},{\bf k}_{-})=0$.
Invoking the approximation $D_L(\Omega,{\bf K})\simeq-(\Omega-v_0 K_z)^2=0$ and $D_A(\omega_{-},{\bf k}_{-}) \simeq
c^2 (k_0-K_z)^2+c^2 K_\perp^2-(\omega_0-\Omega)^2=0$, we obtain $\Omega = v_0 K_z$ and the resonance condition
$(K_z-2\gamma_0^2 k_0)^2+\gamma_0^2 K_\perp^2=4\gamma_0^4 k_0^2$ for $\omega_0 \approx c k_0$ and $v_0\approx -c$. 
The corresponding resonant wave vector components of the radiation field, ${\bf k}_{-}={\bf k}_{0}-{\bf K}$, shown in 
top panels of Fig. \ref{resonant}, form an ellipsoid in wave vector space rotationally symmetric around the $k_{z-}$ axis.
The resulting radiation frequency $\omega_{-}=\omega_0-v_0 K_z \simeq \omega_0+cK_z$ is strongly upshifted in the parallel direction ($K_\perp=0$), where we have $K_z=4\gamma^2 k_0 /\gamma_A^2$ and $\omega_{-}\approx 4\gamma^2 c k_0 /\gamma_A^2$. 
The result differs by a factor two when compared with the case involving a static wiggler \cite{static}.

Comparing the two terms under the square root in Eq. (\ref{Ohm_pm}), we see that the quantum recoil effect starts to 
be important in the parallel direction ($K_\perp=0$, $K_z\approx 4\gamma^2 k_0/\gamma_A^2$) when $k_0 \approx k_{0,crit}$,
where 
\begin{equation}
  k_{0,crit}=\bigg(\frac{\omega_{pe}\gamma_A^3 m_e}{8\hbar \gamma^{5/2}}\bigg)^{1/2}. 
  \label{wcrit}
\end{equation}
In the classical limit $k_0\ll k_{0,crit}$ (corresponding to the Raman regime discussed below), 
we have $\Omega_{p}=\omega_{pe}\gamma_A/\gamma^{3/2}$, while for $k_0 \gg k_{0,crit}$, the quantum 
effects dominate and  we have $\Omega_{p}= \hbar K_z^2 \gamma_A^2/2\gamma^3 m_e 
= 8\hbar \gamma k_0^2/\gamma_A^2 m_e$. An expression analogous to (\ref{wcrit}) can be derived 
for the static wiggler  case \cite{static}.

For $\Omega\approx v_0 K_z$, $K_z\approx 4\gamma^2 k_0/\gamma_A^2$, and $\omega_{0}^2\gg \omega_{pe}^2$,
we have $D_A(\Omega,{\bf K})\approx 4 c^2 K_z k_0$ in Eq. (\ref{coupling}), so that
\begin{equation}
  D_L(\Omega,{\bf K}) D_A(\omega_{-},{\bf k}_{-})= \frac{2 \omega_{pe}^2 c^2 K_z k_0}{\gamma^3} a_0^2.
  \label{coupling2}
\end{equation}
Setting $\Omega=v_0 K_z+\Omega_{p}+i\Gamma$, where the real part $\Gamma_R$ of $\Gamma$ is the growth rate,
and choosing $K_z$ and $\Omega$ so that $D_A(\omega_{-},{\bf k}_{-})=D_L(\Omega,{\bf K})=0$ for $\Gamma=0$, we obtain
$D_L= -2 i \Omega_p \Gamma+ \Gamma^2$ and  $D_A= 2 i \Gamma(\omega_0-v_0 K_z-\Omega_{p})+\Gamma^2
\approx 2 i \Gamma(\omega_0-v_0 K_z) \approx 2 i \Gamma c K_z$ for  $K_z \gg k_0$ and $v_0\approx -c$. 
Hence, inserting the expressions for $D_L$ and $D_A$  into Eq. (\ref{coupling2}), we have
\begin{equation}
  \Gamma^2(2\Omega_{p}+i\Gamma)= \frac{\omega_{pe}^2 c k_0}{\gamma^3} a_0^2.
  \label{coupling3}
\end{equation}

For $|\Gamma|\gg \Omega_{p}$, we are in the Compton regime where the ponderomotive potential of the laser dominates, 
with the growth rate of the instability given by
\begin{equation}
   \Gamma_R=\frac{\sqrt{3}}{2}\frac{(\omega_{pe}^2 c k_0)^{1/3}}{\gamma} a_0^{2/3}.
  \label{Gamma_R}
\end{equation}
For this case, the quantum recoil effect is negligible \cite{Serbeto09}. On the other hand, for $|\Gamma|\ll |\Omega_{p}|$, 
we have an instability with the growth rate
\begin{equation}
  \Gamma_R=\bigg(\frac{\omega_{pe}^2 c k_0}{2 \gamma^3 \Omega_{p}}\bigg)^{1/2} a_0.
  \label{Gamma}
\end{equation}
Clearly, since $\Omega_{p}$ is in the denominator, the quantum recoil effect leads to a decrease of the growth rate. 
Comparing Eqs. (\ref{Gamma_R}) and (\ref{Gamma}), we find that  the limiting amplitude between the two regimes is 
given by $a_0=a_{crit}$, where
\begin{equation}
  a_{crit}=\bigg(\frac{27 \gamma^3 \Omega_{p}^3}{8\omega_{pe}^2 c k_0}\bigg)^{1/2}.
\end{equation}
Equation (\ref{Gamma_R}) is valid for $a_0 \gg a_{crit}$ and Eq. (\ref{Gamma}) for $a_0 \ll a_{crit}$.
In the Raman regime $k_0\ll k_{0,crit}$, Eq.~(\ref{Gamma}) gives the growth rate 
$\Gamma_R \approx (\omega_{pe}c k_0/2 \gamma^{3/2}\gamma_A)^{1/2} a_0$, while in the quantum regime $k_0\gg k_{0,crit}$,
Eq. (\ref{Gamma}) yields $\Gamma_R \approx (\omega_{pe}^2 \gamma_A^2 m_e c/16\gamma^4 \hbar k_0)^{1/2} a_0$.
In Fig. 1, we have illustrated different regimes for the FEL instability, including the quantum and Raman
regimes for small amplitude wiggler fields, and the Compton regime for large amplitudes. The transition from the
quantum to the Compton regime in Fig. 1 corresponds to the quantum FEL parameter \cite{Bonifacio05} ${\bar{\rho}}
=\rho m_e c \gamma_r/\hbar K_z$ going from smaller to larger values than unity, where 
$\rho=(a_0 \omega_{pe}/4 c k_0)/\gamma_r$ is the classical BNP parameter \cite{Bonifacio84}, 
and $\gamma_r$ the resonant energy in $m_e c^2$ units.

\begin{figure}
  \centering
  \includegraphics[width=8.5cm]{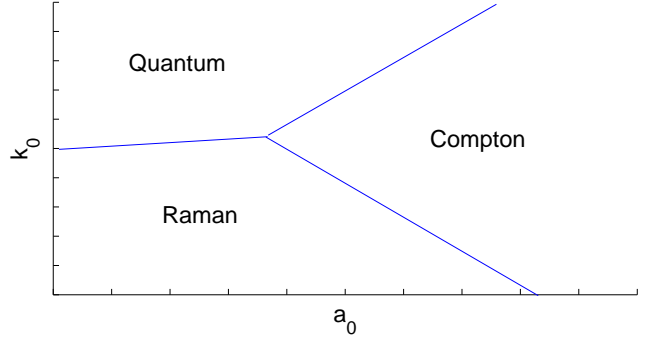}
  \caption{Schematic picture of different regimes for the FEL instability, 
  showing the quantum regime $k_0 > k_{0,crit}$, the Raman 
  regime $k_0 < k_{0,crit}$, and the Compton regime $a_0 > a_{crit}$.}
\end{figure}

\begin{figure}
  \centering
  \includegraphics[width=8.5cm]{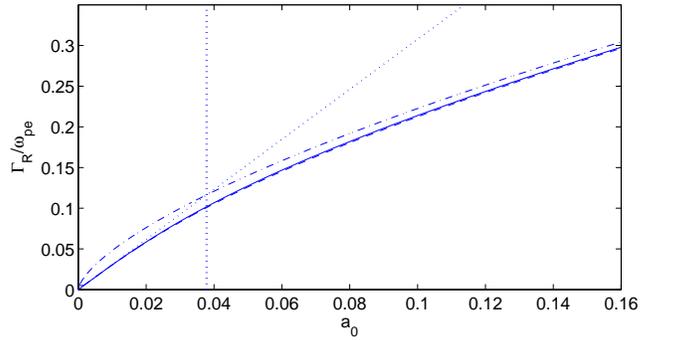}
  \caption{The maximum growth rate as a function of $a_0$ for $K_{\perp}=0$ and $\gamma=5$, 
using the full dispersion relation (\ref{FEL_disp2}) (solid curve),
and the approximations (\ref{coupling3}), (\ref{Gamma_R}) and (\ref{Gamma}) (dashed, dash-dotted and dotted curves, respectively).
The vertical dotted line indicates $a_0=a_{crit}=0.038$.}
  \label{Gamma_p5}
\end{figure}

For illustrative purposes, we choose a beam density $n_0=2.2\times 10^{22}\,\mathrm{m^{-3}}$, 
giving $\omega_{pe}= 8.37\times 10^{12}\,\mathrm{s}^{-1}$, $a_0=0.15$, 
and a wiggler wavelength of $\lambda_0=1\,\mathrm{\mu m}$ giving 
$k_0=2\pi/\lambda_0= 6.36\times 10^{6}\,\mathrm{m}^{-1}$ \cite{Piovella08}. For
$\gamma=5$ one has $k_{0,crit}=1.27\times 10^{7}\,\mathrm{m}^{-1} > k_0$, so that the plasma oscillation effect
dominates over the quantum recoil effect. Figure \ref{Gamma_p5} displays the growth rate as a function of  $a_0$, 
obtained from the exact dispersion relation (\ref{FEL_disp2}) and from the approximations (\ref{coupling3}), 
(\ref{Gamma}) and (\ref{Gamma_R}). We note that the growth rate obtained from (\ref{coupling3})
agrees very well with the one obtained from (\ref{FEL_disp2}). Since $a_{crit}=0.038<a_0$,  the ponderomotive force dominates over the 
plasma and quantum oscillations, so that Eq.~(\ref{Gamma_R}) can be used to calculate the growth rate, giving $\Gamma_R=2.6\times10^{12}\mathrm{s}^{-1}$ and an interaction length scale $c/\Gamma_R\approx 0.1\,\mathrm{mm}$. 

On the other hand, due to the quantum recoil effect, the gain can rapidly decrease for higher values of $\gamma$. 
Using the same parameters as above but $\gamma=36$ \cite{Piovella08}, we have 
$k_{0,crit}= 1.07\times 10^{6} \mathrm{m}^{-1} < k_0$, so that
the quantum recoil effect dominates the beam oscillations. Here we have $a_{crit}= 1.6 \gg a_0$, so that
Eq. (\ref{Gamma}) can be used to estimate the growth rate, which gives $\Gamma_R= 1.5\times 10^{11}\,\mathrm{s}^{-1}$ 
and an interaction length $c/\Gamma_R=2\,\mathrm{mm}$. For this case, the expression (\ref{Gamma_R})
over-estimates the growth rate to $\Gamma_R/\omega_{pe}\approx 3.5\times 10^{11}\mathrm{s}^{-1}$, 
giving $c/\Gamma_R\approx 1\,\mathrm{mm}$. 

\begin{figure}
  \centering
  \includegraphics[width=8.5cm]{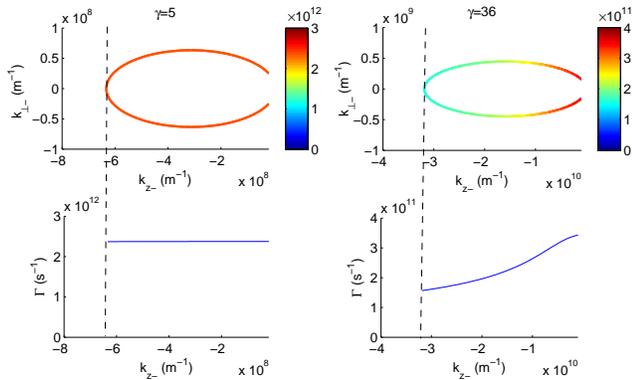}
  \caption{Resonant parallel and perpendicular radiation wavenumbers $k_{z-}$ and $k_{\perp-}$ for 
  $\gamma=5$ and $\gamma=36$ (top panels) with  the corresponding growth rate  $\Gamma\,(\mathrm{s}^{-1})$ 
shown in color. Bottom panels show the growth rate as a  function of $k_{z-}$. (The scalings of the vertical axes 
are enhanced in the top panels.) The vertical bars show the location of the largest resonant wave numbers 
$\approx - 4 k_0\gamma^2/\gamma_A^2$.}
  \label{resonant}
\end{figure}

The instability of oblique scattering is shown in Fig.~3 for resonant radiation wavenumbers $k_{z-}=k_0-K_z$ and $k_{\perp-}=-K_\perp$,
obeying the resonance condition $(K_z-2\gamma_0^2 k_0)^2+\gamma_0^2 K_\perp^2=4\gamma_0^4 k_0^2$ derived above. The growth
rate, deduced from Eq.~(\ref{coupling3}), is almost independent of the radiation wavenumbers for $\gamma=5$, where quantum recoil effects
are unimportant. For $\gamma=36$, there is a significant decrease of the growth rate for larger radiation wavenumbers due to quantum recoil effects, primarily in the parallel direction. For too wide electron beams, it could lead to a broad-band radiation emission due to oblique scattering, 
while for narrow electron beams this is prevented due to a decrease of the possible interaction length in the perpendicular direction.

Summarizing, we have presented a nonlinear model for relativistic quantum X-ray FELs, using a collective  Klein-Gordon model for 
relativistic electrons, coupled with the  Maxwell equations for the EM fields, for an arbitrarily large amplitude laser wiggler field. 
We have derived a nonlinear dispersion relation for the amplification of the radiation due to scattering instability in three-space dimensions. 
It is found that quantum recoil effects can decrease the growth rate of the resonant instability, primarily parallel to the beam direction,
increasing the interaction length over which the radiation amplification occurs. The present study has
assumed that the coherence of the relativistic electron beam and its transverse emittance \cite{Huang07} are unaffected by the 
quantum effects over the scale length of out interest. The quantum effect could be important if the thermal de Broglie
wavelength $\lambda_{th}=\hbar\sqrt{2\pi/m_e k_B T_e}$ is comparable to the inter-particle distance $n_0^{-1/3}$ \cite{Claessens05}. 
For $n_0=2.2\times 10^{22}\,\mathrm{m}^{-3}$ this happens only for $T_e<4\,\mathrm{K}$ when the beam electrons are Fermi-degenerate.
At room temperature and above, the thermal effects clearly dominate over the quantum degeneracy effects on the beam emittance.
On the other hand, quantum diffusion due to spontaneous photon emission could 
lead to an increase of the energy spread of the electron beam. To estimate the relative energy spread, we use the formula \cite{Saldin96}
$\Delta \gamma^2 = (14/15) \lambda_C r_e \gamma^4 k_0^3 a_0^2 F(a_0) \Delta z$, where $F(a_0)\approx 1$
for $a_0<1$, $r_e\approx 2.8\times 10^{-15}\,\mathrm{m}$ is the classical electron radius, and $\Delta z$ is the interaction
distance, which can be taken to be 10 e-foldings, $\Delta z=10\times c/\Gamma_R$. For the case $\gamma=5$ above we obtain
the energy spread $\Delta \gamma = 6\times 10^{-5}$, while for $\gamma=36$ we have the relatively 
large value $\Delta \gamma = 0.014$, which might influence the performance of the FELs.
In conclusion, we stress that our results  will provide a  guideline for designing new experiments 
for generating tunable X-ray FELs by using tenuous relativistic electron beams and intense laser wigglers.

\acknowledgments
This work was supported by the Deutsche Forschungsgemeinschaft (DFG) through the Project SH21/3-2 of the Research Unit 1048.


\begin{thebibliography}{99}
  \bibitem{Hand09} E. Hand, Nature (London) {\bf 461}, 708 (2009).
  \bibitem{Glenzer} S. H. Glenzer {\it et al.}, Phys. Rev. Lett. {\bf 98}, 065002 (2007);
P. Neumayer {\it et al.}, {\it ibid.} {\bf 105}, 075003 (2010); S. H. Glenzer and R. Redmer, Rev. Mod. Phys. {\bf 81}, 1625 (2009).
  \bibitem{Andreev} A. V. Andreev, JETP Lett. {\bf 72}, 238 (2000);G. Mourou {\it et al.}, Rev. Mod. Phys. {\bf 78}, 309 (2006);
P. K. Shukla and B. Eliasson, Rev. Mod. Phys. {\bf 83}, 885 (2011); M. Marklund and P. K. Shukla, {\it ibid.} {\bf 78}, 591 (2006).
  \bibitem{Madey71} J. M. J. Madey, J. Appl. Phys. {\bf 42}, 1906 (1971).
  \bibitem{Hopf76} F. A. Hopf {\it et al.}, Phys. Rev. Lett. {\bf 37}, 1342 (1976);
 T. Kwan {\it et al.}, Phys. Fluids {\bf 20}, 581 (1977); N. M. Kroll and W. A. McMullin, Phys. Rev. A {\bf 17}, 300 (1978).
  \bibitem{Bonifacio85} R. Bonifacio {\it et al.}, Nucl. Instrum. Methods A {\bf 237}, 168 (1985).
  \bibitem{Schroeder01} C. B. Schroeder {\it et al.}, Phys. Rev. E {\bf 64}, 056502 (2001).
  \bibitem{Bonifacio05} R. Bonifacio {\it et al.}, Nucl. Instrum. Methods A {\bf 543}, 645 (2005).
  \bibitem{Bonifacio05b} R. Bonifacio {\it et al.}, Europhys. Lett. {\bf 69}, 55 (2005).
  \bibitem{Yan86} Y. T. Yan and J. M. Dawson, Phys. Rev. Lett. {\bf 57}, 1599 (1986);  C. Joshi  {\it et al.}, 
IEEE J. Quantum Electron. {\bf QE-23}, 1571 (1987).
  \bibitem{McIver79} J. K. McIver and M. V. Federov, Sov. Phys. JETP {\bf 49}, 1012 (1979); I. V. Smetanin, Laser Phys. {\bf 7}, 318 (1997).
  \bibitem{Preparata88} G. Preparata, Phys. Rev. A {\bf 38}, 233 (1988).
  \bibitem{Bonifacio94} R. Bonifacio {\it et al.}, Phys. Rev. Lett. {\bf 73}, 70 (1994).
  \bibitem{Bonifacio08} R. Bonifacio {\it et al.}, Nucl. Instrum. Methods Phys. Res. A {\bf 593}, 69 (2008).
  \bibitem{Serbeto08} A. Serbeto {\it et al.}, Phys. Plasmas {\bf 15}, 013110 (2008).
  \bibitem{Piovella08} N. Piovella, {\it et al.}, Phys. Rev. Lett. {\bf 100}, 044801 (2008); M. M. Cola, {\it et al.}, Nucl. Instrum. Methods 
Phys. Res. A {\bf 593}, 75 (2008).
  \bibitem{Serbeto09} A. Serbeto, L. F. Monteiro, K. H. Tsui, and J. T. Mendon\c{c}a, Plasma Phys. Control. Fusion {\bf 51} 124024 (2009).
  \bibitem{Takabayasi53} T. Takabayasi, Prog. Theor. Phys. {\bf 9}, 187 (1953).
  \bibitem{Eliasson11} B. Eliasson and P. K. Shukla, Phys. Rev. E {\bf 83}, 046407 (2011).
  \bibitem{Drake74} J. F. Drake {\it et al.}, Phys. Fluids {\bf 17}, 778 (1974).
  \bibitem{static} For a static wiggler \cite{Schroeder01} we would have $\omega_0=0$ in the expression for $D_A$, 
    with the result $K_z=2 k_0 \gamma^2/\gamma_A^2$ and $\omega_-=-\Omega=2\gamma^2 c k_0/\gamma_A^2$, 
which differs by a factor two from the electromagnetic wiggler. Using $K_z=2 \gamma_0^2 k_0 /\gamma_A^2$ in Eq. (\ref{Ohm_pm}), 
we obtain, analogously to (\ref{wcrit}), $k_{0,crit}=(\omega_{pe}\gamma_A^3 m_e/2 \gamma^{5/2}\hbar)^{1/2}$, with
$\Omega_{p}=\omega_{pe}\gamma_A/\gamma^{3/2}$ for $k_0\ll k_{0,crit}$ and $\Omega_{p}=2\hbar \gamma k_0^2 / \gamma_A^2 m_e$ 
for $k_0\gg k_{0,crit}$.
  \bibitem{Bonifacio84} R. Bonifacio {\it et al.}, Opt. Commun. {\bf 50}, 373 (1984).
  \bibitem{Huang07}  Z. Huang and K.-J. Kim, Phys. Rev. ST Accel. Beams {\bf 10}, 034801 (2007).
  \bibitem{Claessens05} B. J. Claessens {\it et al.}, Phys. Rev. Lett. {\bf 95}, 164801 (2005).
  \bibitem{Saldin96} 	
	E. L. Saldin, E. A. Schneidmiller, and M. V. Yurkov, 
	Nucl. Instr. Meth. Phys. Res. Sect. A {\bf 381}, 545 (1996).
\end{thebibliography}
\end{document}